\documentclass[10pt,a4paper,twocolumn]{elsart}
\usepackage{natbib}

\usepackage{epsfig}
\usepackage{amsfonts}
\usepackage{amsmath}
\usepackage{axodraw}

\epsfverbosetrue

\def\3{\ss}

\newcommand{\tev}{{\rm Te}\kern-1.pt{\rm V}}
\newcommand{\gev}{{\rm Ge}\kern-1.pt{\rm V}}
\newcommand{\mev}{{\rm Me}\kern-1.pt{\rm V}}
\newcommand{\kev}{{\rm Ke}\kern-1.pt{\rm V}}
\newcommand{\gevsq}{\mbox{$\mathrm{{\rm Ge}\kern-1.pt{\rm V}}^2$}}
\newcommand{\gevmsq}{\mbox{$\mathrm{{\rm Ge}\kern-1.pt{\rm V}}^{-2}$}}

\begin{document}
\runauthor{Cranmer and Bowman}
\begin{frontmatter}
\title{PhysicsGP: A Genetic Programming Approach to %
 Event Selection}

\author[University of Wisconsin-Madison]{Kyle Cranmer}
\author[University of Arkansas at Fayetteville]{R. Sean Bowman}

\address[University of Wisconsin-Madison]{CERN, CH-1211 Geveva, Switzerland}
\address[University of Arkansas at Fayetteville]{Open Software
Services, LLC, Little Rock, Arkansas, USA}
\begin{abstract}
We present a novel multivariate classification technique based on
Genetic Programming.  The technique is distinct from Genetic
Algorithms and offers several advantages compared to Neural Networks
and Support Vector Machines.  The technique optimizes a set of
human-readable classifiers with respect to some user-defined
performance measure.
We calculate the Vapnik-Chervonenkis dimension of this class of
learning machines and consider a practical example: the search for the
Standard Model Higgs Boson at the LHC.  The resulting classifier is
very fast to evaluate, human-readable, and easily portable.  The
software may be downloaded at:
\texttt{http://cern.ch/$\sim$cranmer/PhysicsGP.html}

\end{abstract}
\begin{keyword}
Genetic Programming, Triggering, Classification, VC Dimension, Genetic Algorithms, Neural Networks, Support Vector Machines
\end{keyword}
\end{frontmatter}

\pagenumbering{arabic}

\section{Introduction}

The use of multivariate algorithms in the search for particles in High
Energy Physics has become quite common.  Traditionally, a search can
be viewed from a classification point of view: from a tuple of
physical measurements (i.e., momenta, energy, etc.) we wish to
classify an event as signal or background.  Typically, this
classification is realized through a Boolean expression or {\it cut}
designed by hand.  The high dimensionality of the data makes this
problem difficult in general and favors more sophisticated
multivariate algorithms such as Neural Networks, Fisher Discriminants,
Kernel Estimation Techniques, or Support Vector Machines.  This paper
focuses on a Genetic Programming approach and considers a specific
example: the search for the Higgs Boson at the LHC.

The use of Genetic Programming for classification is fairly limited;
however, it can be traced to the early works on the subject by
Koza~\cite{koza:gp}.  More recently, Kishore \textit{et al}.\ extended
Koza's work to the multicategory problem~\cite{kishore:2000}.  To the
best of the authors' knowledge, the work presented in this paper is
the first use of Genetic Programming within High Energy Physics.

In Section~\ref{S:History} we provide a brief history of evolutionary
computation and distinguish between Genetic Algorithms (GAs) and
Genetic Programming (GP).  We describe our algorithm in detail for an
abstract performance measure in Section~\ref{S:PhysicsGP} and discuss
several specific performance measures in
Section~\ref{S:PerformanceMeasures}.

Close attention is paid to the performance measure in order to
leverage recent work applying the various results of statistical
learning theory in the context of new particle searches.  This recent
work consists of two components.  In the first, the Neyman-Pearson
theory is translated into the {\it Risk}
formalism~\cite{Cranmer:Acta,Cranmer:2003vu}.  The second component
requires calculating the Vapnik-Chervonenkis dimension for the
learning machine of interest.  In Section~\ref{S:Vapnik}, we calculate
the Vapnik-Chervonenkis dimension for our Genetic Programming
approach.

Because evolution is an operation on a population, GP has some
statistical considerations not found in other multivariate algorithms.
In Section~\ref{S:Fluctuations} we consider the main statistical
issues and present some guiding principles for population size based
on the user-defined performance measure.

Finally, in Section~\ref{S:Higgs} we examine the application of our
algorithm to the search for the Higgs Boson at the LHC.  

\section{Evolutionary Computation}\label{S:History}

In Genetic Programming (GP), a group of ``individuals'' evolve and
compete against each other with respect to some performance measure.
The individuals represent potential solutions to the problem at hand,
and evolution is the mechanism by which the algorithm optimizes the
population.  The performance measure is   a
mapping that assigns a fitness value to each individual.  GP can be
thought of as a Monte Carlo sampling of a very high dimensional search
space, where the sampling is related to the fitness evaluated in the
previous generation.  The sampling is not ergodic -- each generation
is related to the previous generations -- and intrinsically takes
advantage of stochastic perturbations to avoid local extrema%
\footnote{These are the properties that give power to Markov Chain
Monte Carlo techniques.}.

Genetic Programming is similar to, but distinct from Genetic
Algorithms (GAs), though both methods are based on a similar
evolutionary metaphor.  GAs evolve a bit string which typically
encodes parameters to a pre-existing program, function, or class of
cuts, while GP directly evolves the programs or functions.  For
example, Field and Kanev~\cite{Field:1997kt} used Genetic Algorithms
to optimize the lower- and upper-bounds for six 1-dimensional cuts on
Modified Fox-Wolfram ``shape'' variables.  In that case, the
phase-space region was a pre-defined 6-cube and the GA was simply
evolving the parameters for the upper and lower bounds.  On the other
hand, our algorithm is not constrained to a pre-defined shape or
parametric form.  Instead, our GP approach is concerned directly with
the construction and optimization of a nontrivial phase space region
with respect to some user-defined performance measure.

In this framework, particular attention is given to the performance
measure.  The primary interest in the search for a new particle is
hypothesis testing, and the most relevant measures of
performance are the expected statistical significance (usually
reported in Gaussian sigmas) or limit setting potential.  The
different performance measures will be discussed in
Section~\ref{S:PerformanceMeasures}, but consider a concrete example:
$s/\sqrt{b}$, where $s$ and $b$ are the number of signal and
background events satisfying the event selection, respectively.


\section{The Genetic Programming Approach}\label{S:PhysicsGP}

\begin{figure}
\begin{center}
\epsfig{file=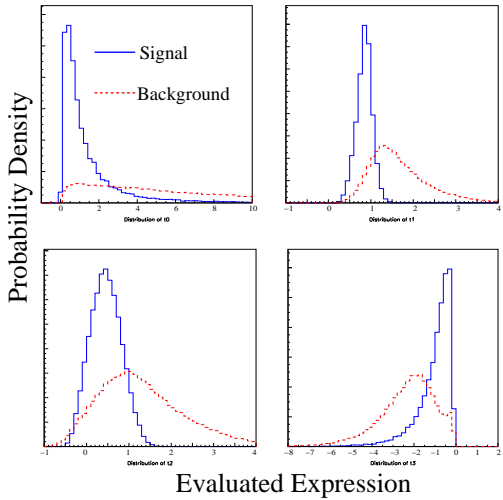, height= 2.6 in}
\caption{Signal and Background histograms for an expression.  }
\label{fig:truthDistributions}
\end{center}
\end{figure}

While the literature is replete with uses of Genetic Programming and
Genetic Algorithms, direct evolution of cuts appears to be novel.  In
the case at hand, the individuals are composed of simple arithmetic
expressions, $f$, on the input variables $\vec{v}$.  Without loss of
generality, the cuts are always of the form $-1 < f(\vec{v}) < 1$.  By
scaling, $f(\vec{v})\rightarrow a f(\vec{v})$, and translation,
$f(\vec{v})\rightarrow f(\vec{v}) + b$, of these expressions, single-
and double-sided cuts can be produced.  An individual may consist of
one or more such cuts combined by the Boolean conjunction {\tt AND}.
Fig.~\ref{fig:truthDistributions} shows the signal and background
distributions of four expressions that make up the most fit individual
in a development trial.

Due to computational considerations, several structural changes have
been made to the na\"\i ve implementation.  First, an Island Model of
parallelization has been implemented (see Section~\ref{S:Island}).
Secondly, individuals' fitness can be evaluated on a randomly chosen
sub-sample of the training data, thus reducing the computational
requirements at the cost of statistical variability.  There are
several statistical considerations which are discussed in
Section~\ref{S:Fluctuations}.

    \subsection{Individual Structure, Mutation, and Crossover}

\begin{figure}[Ht]
\begin{center}
\epsfig{file=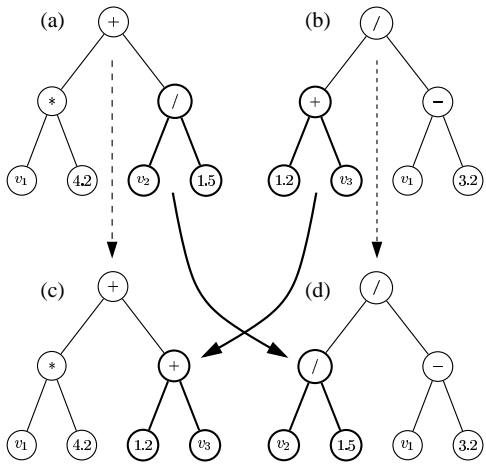, height= 2.4 in}
\caption{An example of crossover.  At some given generation, two
parents (a) and (b) are chosen for a crossover mutation.  Two
subtrees, shown in bold, are selected at random from the parents and
are swapped to produce two children (c) and (d) in the subsequent
generation.}
\label{fig:crossover}
\end{center}
\end{figure}

The genotype of an individual is a collection of expression trees
similar to abstract syntax trees that might be generated by a compiler
as an intermediate representation of a computer program.  An example
of such a tree is shown in Fig.~\ref{fig:crossover}a which corresponds
to a cut $| 4.2v_1 + v_2/1.5 |< 1$.  Leafs are either constants or one
of the input variables.  Nodes are simple arithmetic operators:
addition, subtraction, multiplication, and safe division%
\footnote{Safe division is used to avoid division by zero.}.%
When an individual is presented with an event, each expression
tree is evaluated to produce a number.  If all these numbers lie
within the range $(-1,1)$, the event is considered signal.
Otherwise the event is classified as background.

Initial trees are built using the PTC1 algorithm described
in~\cite{luke00two}.  After each generation, the trees are modified by
\textit{mutation} and \textit{crossover}.  Mutation comes in two
flavors.  In the first, a randomly chosen expression in an individual
is scaled or translated by a random amount.  In the second kind of
mutation, a randomly chosen subtree of a randomly chosen expression is
replaced with a randomly generated expression tree using the same
algorithm that is used to build the initial trees.

While mutation plays an important r\^{o}le in maintaining genetic
diversity in the population, most new individuals in a particular
generation result from crossover.  The crossover operation takes two
individuals, selects a random subtree from a random expression from
each, and exchanges the two.  This process is illustrated in
Fig.~\ref{fig:crossover}. 

	\subsection{Recentering}

Some expression trees, having been generated randomly, may prove to be
useless since the range of their expressions over the domain of
their inputs lies well outside the interval $(-1,1)$ for every input
event.  When an individual classifies all events in the same way
(signal or background), each of its expressions is translated to the
origin for some randomly chosen event exemplar $\vec{v_0}$, {\it viz.}
$f(\vec{v})~\rightarrow~f(\vec{v})~-~f(\vec{v_0})$.  This modification
is similar to, and thus reduces the need for, normalizing input variables.

	\subsection{Fitness Evaluation}

Fitness evaluation consumes the majority of time in the execution of
the algorithm.  So, for speed, the fitness evaluation is done in C.
Each individual is capable of expressing itself as a fragment of C
code.  These fragments are pieced together by the Python program,
written to a file, and compiled.  After linking with the training
vectors, the program is run and the results communicated back to the
Python program using standard output.  

The component that serializes the population to C and reads the
results back from the generated C program is configurable, so
that a user-defined performance measure may be implemented.

	\subsection{Evolution \& Selection Pressure}

\begin{figure}
\begin{center}
\epsfig{file=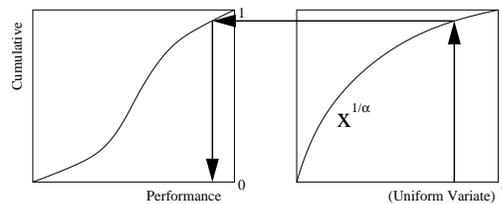, width= 2.6in}
\caption{Monte Carlo sampling of individuals based on their fitness.
A uniform variate $x$ is transformed by a simple power to produce
selection pressure: a bias toward individuals with higher fitness.}
\label{fig:SelectionPressure}
\end{center}
\end{figure}

After a given generation of individuals has been constructed and the
individuals' fitnesses evaluated, a new generation must be constructed.
Some individuals survive into the new generation, and some new
individuals are created by mutation or crossover.  In both cases, the
population must be sampled randomly.  To mimic evolution, some {\it
selection pressure} must be placed on the individuals for them to
improve.  This selection pressure is implemented with a simple Monte
Carlo algorithm and controlled by a  parameter $\alpha >1$.  The
procedure is illustrated in Fig.~\ref{fig:SelectionPressure}.  In a
standard Monte Carlo algorithm, a uniform variate $x\in [0,1]$ is
generated and transformed into the variable of interest by the
inverse of  its cumulative distribution.  Using the cumulative
distribution of the fitness will exactly reproduce the population
without selection pressure; however, this sampling can be biased with
a simple transformation.  The right plot of
Fig.~\ref{fig:SelectionPressure} shows a uniform variate $x$ being
transformed into $x^{1/\alpha}$, which is then inverted (left plot) to
select an individual with a given fitness.  As the parameter $\alpha$
grows, the individuals with high fitness are selected increasingly
often.

\begin{figure}
\epsfig{file=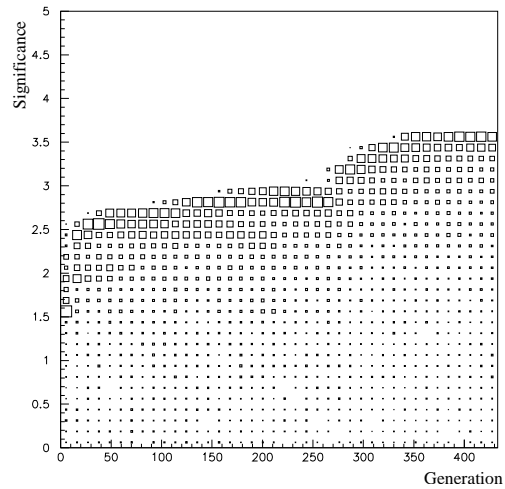, height= 2.5 in}
\caption{The fitness of the population as a function of time.  This
plot is analogous to a neural network error vs. epoch plot, with the
notable exception that it describes a population and not an
individual.  In particular, the neural network graph is a 1-dimensional
curve, but this is a two dimensional distribution.}
\label{fig:evolution}
\end{figure}

While the selection pressure mechanism helps the system evolve, it
comes at the expense of genetic diversity.  If the selection pressure
is too high, the population will quickly converge on the most fit
individual.  The lack of genetic diversity slows  evolutionary
progress.  This behavior can be identified easily by looking at plots
such as Fig.~\ref{fig:evolution}.  We have found that a moderate
selection pressure $\alpha \in [1,3]$ has the best results.

	\subsection{Parallelization and  the Island Model}\label{S:Island}

GP is highly concurrent, since different individuals' fitness
evaluations are unrelated to each other, and dividing the total
population into a number of sub-populations is a simple way to
parallelize a GP problem.  Even though this is a trivial modification
to the program, it has been shown that such coarse grained
parallelization can yield greater-than-linear
speedup~\cite{david95parallel}.  Our system uses a number of Islands
connected to a Monitor in a star topology.  CORBA is used to allow the
Islands, which are distributed over multiple processors, to
communicate with the Monitor.

Islands use the Monitor to exchange particularly fit individuals each
generation.  Since a separate monitor process exists, a synchronous
exchange of individuals is not necessary.  The islands are virtually
connected to each other (via the Monitor) in a ring topology.  

\section{Performance Measures}\label{S:PerformanceMeasures}

The Genetic Programming approach outlined in the previous section is a
very general algorithm for producing individuals with high fitness,
and it allows one to factorize the definition of fitness from the
algorithm.  In this section we examine the function(al) which assigns
each individual its fitness: the performance measure.

Before proceeding, it is worthwhile to compare GP to popular
multivariate algorithms such as Support Vector Machines and Neural
Networks.  Support Vector Machines typically try to minimize the risk
of misclassification $\sum_i |y_i - f(\vec{v}_i)|$, where $y_i$ is the
target output (usually 0 or -1 for background and 1 for signal) and
$f(\vec{v}_i)$ is the classification of the $i^{th}$ input.  This is
slightly different than the error function that most Neural Networks
with backpropagation attempt to minimize: $\sum_i |y_i -
f(\vec{v}_i)|^2$~\cite{Werbos,PDP1}.  In both cases, this performance
measure is usually hard-coded into a highly optimized algorithm and
cannot be easily replaced.  Furthermore, these two choices are not
always the most appropriate for High Energy Physics, as will be
discussed in Section~\ref{SS:Direct}.

The most common performance measure for a particle search is the
Gaussian significance, $s/\sqrt{b}$, which measures the statistical
significance (in ``sigmas'') of the presence of a new signal.  The
performance measure $s/\sqrt{b}$ is calculated by determining how many
signal events, $s$, and background events, $b$, a given individual
will select in a given amount of data (usually measured in
fb$^{-1}$).

The $s/\sqrt{b}$ is actually an approximation of the Poisson
significance, $\sigma_P$, the probability that an expected background
rate $b$ will fluctuate to $s+b$.  The key difference between the two
is that as $s,b\rightarrow 0$, the Poisson significance will always
approach 0, but the Gaussian significance may diverge.  Hence, the
Gaussian significance may lead to highly fit individuals that accept
almost no signal or background events.

The next level of sophistication in significance calculation is to
include systematic error in the background only prediction $b$.  These
calculations tend to be more difficult and the field has not adopted a
standard~\cite{Cousins:1992qz,Cranmer:2003vt, Linnemann:2003vw}.  It
is also quite common to improve the statistical significance of an
analysis by including a discriminating
variable~\cite{FinalLHWG:2003sz}.

In contrast, one may be more interested in excluding some proposed
particle.  In that case, one may wish to optimize the exclusion
potential.  The exclusion potential and discovery potential of a
search are related, and G. Punzi has suggested a performance measure
which takes this into account quite naturally~\cite{Punzi}.

Ideally, one would use as a performance measure the same procedure
that will be used to quote the results of the experiment.  For
instance, there is no reason (other than speed) that one could not
include discriminating variables and systematic error in the
optimization procedure (in fact, the authors have done both).

\subsection{Direct vs. Indirect Methods}\label{SS:Direct}

Certain approaches to multivariate analysis leverage the many powerful
theorems of statistics, assuming one can explicitly refer to the joint
probability density of the input variables and target values
$p(\vec{v},y)$.  This dependence places a great deal of stress on the
asymptotic ability to estimate $p(\vec{v},y)$ from a finite set of samples
$\{(\vec{v},y)_i\}$.  There are many such techniques for estimating a
multivariate density function $p(\vec{v},y)$ given the
samples~\cite{Scott,Cranmer:2000du}.  Unfortunately, for high
dimensional domains, the number of samples needed to enjoy the
asymptotic properties grows very rapidly; this is known as the {\it
curse of dimensionality}.

Formally, the statistical goal of a new particle search is to 
minimize the rate of Type II error.  This is logically distinct
from, but asymptotically equivalent to, approximating the likelihood
ratio.  In the case of no interference between the signal and
background, this is logically distinct from, but asymptotically
equivalent to, approximating the signal-to-background ratio.  In fact,
most multivariate algorithms are concerned with approximating an
auxiliary function that is one-to-one with the likelihood ratio.
Because the methods are not directly concerned with minimizing the
rate of Type II error, they should be considered {\it indirect
methods}.  Furthermore, the asymptotic equivalence breaks down in most
applications, and the indirect methods are no longer optimal.  Neural
Networks, Kernel Estimation techniques, and Support Vector Machines
all represent indirect solutions to the search for new particles.  The
Genetic Programming approach is a {\it direct method} concerned with
optimizing a user-defined performance measure.

\section{Statistical Learning Theory}\label{S:Vapnik}

In 1979, Vapnik provided a remarkable family of bounds relating the
performance of a learning machine and its generalization {\it
capacity}~\cite{Vapnik:1979}.  The capacity, or Vapnik-Chervonenkis
dimension (VCD) is a property of a set of functions, or learning
machines, $\{f(\vec{v};\alpha)\}$, where $\alpha$ is a set of
parameters for the learning machine~\cite{Vapnik:1968}.  

\begin{figure}\label{fig:VCdim}
\centerline{\epsfig{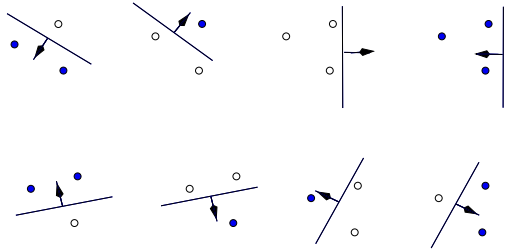}}
\caption{The VCD for a line in $\mathbb{R}^2$ is 3.}
\end{figure}

In the two-class pattern recognition case considered in this paper, an
event $x$ is classified by a learning machine such that
$f(\vec{v};\alpha) \in \{{\rm signal}, {\rm background} \}$.  Given a set of $l$ events each
represented by $\vec{v}_i$, there are $2^l$ possible permutations of
them belonging to the class signal or background.  If for each
permutation there exists a member of the set $\{f(\vec{v};\alpha)\}$
which correctly classifies each event, then we say the set of points
is {\it shattered} by that set of functions.  The VCD for a set of
functions $\{f(\vec{v};\alpha)\}$ is defined as the maximum number of
points which can be shattered by $\{f(\vec{v};\alpha)\}$.  If the VCD
is $h$, it does not mean that every set of $h$ points can be
shattered, but that there exists some set of $h$ points which can be
shattered.  For example, a hyperplane in $\mathbb{R}^n$ can shatter
$n+1$ points (see Fig.~\ref{fig:VCdim} for $n=2$).

In the modern theory of machine learning, the performance of a
learning machine is usually cast in the more pessimistic setting of
{\it risk}.  In general, the risk, $R$, of a learning machine is written as
\begin{equation}\label{E:risk}
R(\alpha) = \int Q(\vec{v},y; \alpha) ~p(\vec{v},y)\,d\vec{v}dy
\end{equation}
where $Q$ measures some notion of {\it loss} between $f(\vec{v}; \alpha)$ and
the target value $y$.  For example, when classifying events, the risk
of mis-classification is given by Eq.~\ref{E:risk} with $Q(\vec{v},y;\alpha)
= |y - f(\vec{v}; \alpha)|$.  Similarly, for regression tasks one takes
$Q(\vec{v},y;\alpha) = (y - f(\vec{v}; \alpha))^2$.  Most of the classic
applications of learning machines can be cast into this formalism;
however, searches for new particles place some strain on the notion of
risk~\cite{Cranmer:Acta,Cranmer:2003vu}.

The starting point for statistical learning theory is to accept that
we might not know $p(\vec{v},y)$ in any analytic or numerical form.  This
is, indeed, the case for particle physics, because only $\{(\vec{v},y)_i\}$
can be obtained from the Monte Carlo convolution of a well-known
theoretical prediction and complex numerical description of the
detector.  In this case, the learning problem is based entirely on the
training samples $\{(\vec{v},y)_i\}$ with $l$ elements.  The risk functional
is thus replaced by the {\it empirical risk functional}
\begin{equation}\label{E:Remp}
R_{\rm emp}(\alpha) = \frac{1}{l} \sum_{i=1}^l Q(\vec{v}_i, y_i; \alpha).
\end{equation}

There is a surprising result that the true risk $R(\alpha)$ can be
bounded independent of the distribution $p(\vec{v},y)$.  In particular, for
$0\le Q(\vec{v},y;\alpha) \le 1$
\begin{eqnarray}\label{E:VCconfidence}
&R(\alpha) &{\le} {R_{\rm emp}(\alpha)}\\ \nonumber
&+&\sqrt{\left(\frac{h(\log(2l/h)+1) - \log(\eta/4)}{l}\right)},
\end{eqnarray}
where $h$ is the VC dimension and $\eta$ is the
probability that the bound is violated.  As $\eta \rightarrow 0$, $h
\rightarrow \infty$, or $l\rightarrow 0$ the bound becomes trivial.
The second term of the right hand side is often referred to as the VC
confidence -- for $h=200$, $l=10^5$, and $\eta=95\%$ the VC confidence
is about 12\%.

While the existence of the bounds found in Eq.~\ref{E:VCconfidence}
are impressive, they are frequently irrelevant.  In particular, for
Support Vector Machines with radial basis functions for kernels the
VCD is formally infinite and there is no bound on the true risk.
Similarly, for Support Vector Machines with polynomial kernels of
degree $p$ and data embedded in $d$ dimensions, the VCD is
$\binom{p+d-1}{p}+1$ which grows very quickly.

This motivates a calculation of the VCD of the GP approach.
 
\subsection{VCD for Genetic Programming}

\begin{figure*}
\begin{alignat}{3} \nonumber
f(x,y; \alpha)  &= a_1  & +a_2\cdot x \qquad  &\qquad +a_3\cdot y \\ \nonumber
 &+a_4\cdot x\cdot x  &\qquad +a_5\cdot x\cdot y \quad &\qquad +a_6\cdot  y\cdot y \\ \nonumber
 &+a_7\cdot x\cdot x\cdot y  &\qquad +a_8 \cdot x \cdot y \cdot y  
 &\qquad +a_9 \cdot x \cdot  x \cdot y \cdot y 
\end{alignat}
\caption{An explicit example of the largest polynomial on two
variables with degree two.  In total, 53 nodes are necessary for this
expression which has only 9 independent parameters.}
\label{fig:combinatorics}
\end{figure*}

The VC dimension, $h$, is a property of a fully specified learning
machine.  It is meaningless to calculate the VCD for GP in general;
however, it is sensible if we pick a particular genotype.  For the
slightly simplified genotype which only uses the binary operations of
addition, subtraction, and multiplication, all expressions are
polynomials on the input variables.  It has been shown that for
learning machines which form a vector space over their parameters,%
\footnote{A learning machine, $\mathcal{F}$, is a vector space if for
any two functions $f,g \in \mathcal{F}$ and real numbers $a,b$ the
function $af+bg \in \mathcal{F}$.  Polynomials satisfy these
conditions.}~%
the VCD is given by the dimensionality of the span of their
parameters~\cite{Sontag}.  Because the Genetic Programming approach
mentioned is actually a conjunction of many such cuts, one also must
use the theorem that the VCD for Boolean conjunctions, $b$, among learning
machines is given by ${\rm VCD}(b(f_1,\dots, f_k)) \le c_k \max_i {\rm
VCD}(f_i)$, where $c_k$ is a constant~\cite{Sontag}.

If we placed no bound on the size of the program, arbitrarily large
polynomials could be formed and the VCD would be infinite.  However,
by placing a bound on either the size of the program or the degree of
the polynomial, we can calculate a sensible VCD.  The remaining step
necessary to calculate the VCD of the polynomial Genetic Programming
approach is a combinatorial problem: for programs of length $L$, what
is the maximum number of linearly independent polynomial coefficients?
Fig.~\ref{fig:combinatorics} illustrates that the smallest program
with nine linearly independent coefficients requires eight additions,
eighteen multiplications, eighteen variable leafs, and nine constant
leafs for a total of 53 nodes.  A small Python script was written to
generalize this calculation.

The Genetic Programming approach with polynomial expressions has a
relatively small VCDs (in our tests with seven variables nothing
more than $h=100$ was found) which affords the relevance of the
upper-bound proposed by Vapnik.

\subsection{VCD of Neural Networks}

In order to apply Eq.~\ref{E:VCconfidence}, one must determine the VC
dimension of Neural Networks.  This is a difficult problem in
combinatorics and geometry aided by algebraic techniques.  Eduardo
Sontag has an excellent review of these techniques and shows that the
VCD of neural networks can, thus far, only be bounded fairly
weakly~\cite{Sontag}.  In particular, if we define $\rho$ as the
number of weights and biases in the network, then the best bounds are
$\rho^2 < h < \rho^4$.  In a typical particle physics neural network
one can expect $100<\rho<1000$, which translates into a VCD
as high as $10^{12}$, which implies $l>10^{13}$ for reasonable bounds
on the risk.  These bounds imply enormous numbers of training samples
when compared to a typical training sample of $10^5$.  Sontag goes on
to show that these shattered sets are incredibly special and that the
set of all shattered sets of cardinality greater than $\mu = 2\rho +
1$ is measure zero in general.  Thus, perhaps a more relevant notion
of the VCD of a neural network is given by $\mu$.

\section{Statistical Fluctuations in the Fitness Evaluation}\label{S:Fluctuations}

In this section we examine the trade-off between the time necessary to
evaluate the fitness of an individual and the accuracy of the
fitness when evaluated on a finite sample of events.
Holding computing resources fixed, the two limiting cases are:
\begin{itemize}
  \item[1] With very large sample sizes, one can expect excellent
    estimation of the fitness of individuals and a clear ``winner'' at
    the expense of very little mutation and poorly optimized individuals.

  \item[2] With very small sample sizes, one can expect many mutations
  leading to individuals with very high fitness which do not
  perform as reported on larger samples.
\end{itemize}

\begin{figure}
\begin{center}
\epsfig{file=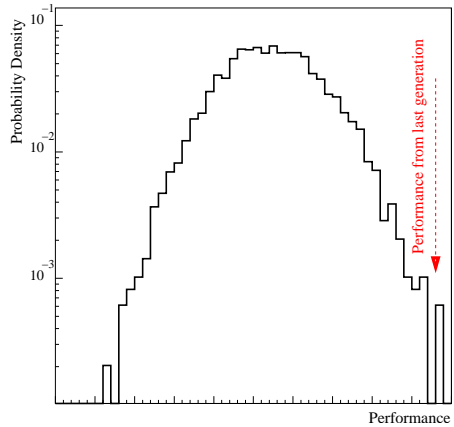, height= 2.2 in}
\end{center}
\caption{The distribution of fitness evaluated on a single individual
with an ensemble of testing samples (each with 400 events).  The
dashed vertical arrow indicates the fitness evaluated with a specific
testing sample in the last generation of training.  The large ($\approx 3\sigma$) upward fluctuation in the evaluated fitness largely
enhanced the chance this individual would be chosen as the winner.}
\label{fig:fitness_prob}
\end{figure}

Illustrated in Fig.~\ref{fig:fitness_prob} is the distribution of
fitness for a given ``winning'' individual generated with a large
ensemble of training samples each with 400 events.  The fitness
reported in the last generation of the training phase (indicated with
an arrow) is much higher than the mean of this distribution.  In fact,
the probability to have randomly chosen a sample of 400 events which
would produce such a high empirical significance is about 0.1\%.

While the chance that an arbitrary individual's fitness evaluates
several standard deviations from the mean is quite small, with
thousands, maybe millions, of individual programs the chance that one
will fluctuate significantly can be quite large.  Furthermore, the
winning individual has a much higher chance of a significant upward
fluctuation, because individuals with upward fluctuations have a higher
chance of being the winner. 

Having recognized that statistical fluctuations in the fitness
evaluation complicate matters, we have developed a few guiding
principles for reliable use of the algorithm.
\begin{itemize}
  \item For training, the standard deviation of the fitness distribution
  evaluated on $N$ events should be on the order of a noticeable and
  marginal improvement in the fitness based on the users performance
  measure.
  \item Select the winning individual with a large testing sample.
\end{itemize}

If we take as our measure of performance $s/\sqrt b$, then it is
possible to calculate the variance due to the fluctuations in $s$ and
$b$.  The expected error is given by the standard propagation of
errors formula.  In particular,
\begin{eqnarray}
\left( \Delta \frac{s}{\sqrt{b}} \right)^2 
&=& \frac{\Delta s^2}{b}  + \frac{\Delta b^2 s^2}{4b^3}  
\end{eqnarray}
where $s=L \sigma_s \epsilon_s$ and $b=L \sigma_b \epsilon_b$ via
standard rate calculations and $\Delta s^2 = \epsilon_s (1-\epsilon_s)
L^2 \sigma_s^2 / N_s$ and similarly $\Delta b^2 = \epsilon_b
(1-\epsilon_b) L^2 \sigma_b^2 / N_b$ via binomial statistics.  For the
analysis presented in Section~\ref{S:Higgs}, the selection efficiency
for signal and background are $\epsilon_s \approx 50\%$ and
$\epsilon_b \approx 5\%$, respectively.  The predicted rate of signal
and background events are $L\sigma_s \approx 100$, and $L\sigma_b
\approx 560$, respectively.  Using these values, one can expect a 10\%
(5\%) relative error on the fitness evaluation with a sample of $N_s=N_b=100$
(400) events.  Analogous calculations can be made for any performance
measure (though they may require numerical studies) to determine a
reasonable sample size.  The rule of thumb that relative errors scale
as $1/\sqrt{N}$ is probably reasonable in most cases.

\section{Case Study: The Search for the Higgs Boson at the LHC}\label{S:Higgs}

Finally we consider a practical example: the search for the Higgs
boson at the LHC.  While there are many channels available, the recent
Vector Boson Fusion analyses offer a sufficiently complicated final
state to warrant the use of multivariate algorithms.

\begin{figure}
\begin{center}
\begin{picture}(175,100)(0,0)
\ArrowLine(10,90)(120,90)\Text(100,95)[b]{$q$}
\ArrowLine(10,10)(120,10)\Text(100,5)[t]{$q$}

\SetColor{Black}
\Photon(50,90)(80,50){-3}{6}	\Text(55,65)[c]{$W$}
\Photon(50,10)(80,50){3}{6}	\Text(55,35)[c]{$W$}
\DashLine(80,50)(120,50){5}  	\Text(100,55)[b]{$H$}
\Photon(120,50)(150,70){3}{6}	\Text(135,70)[r]{$W^+$}
\Photon(120,50)(150,30){-3}{6} 	\Text(135,30)[r]{$W^-$}
\ArrowLine(150,70)(160,90)      \Text(155,90)[b]{$\nu$}
\ArrowLine(170,70)(150,70)      \Text(175,70)[l]{$l^+$}
\ArrowLine(150,30)(170,30)      \Text(175,30)[l]{$l^-$}
\ArrowLine(160,10)(150,30)      \Text(155,10)[t]{$\bar{\nu}$}
\end{picture}
\\
\end{center}
\caption{Tree-level diagram of Vector Boson Fusion Higgs production with $H\rightarrow
W^+W^-\rightarrow l^+l^-\nu\overline{\nu}$}
\label{fig:vbf_feynman}
\end{figure}
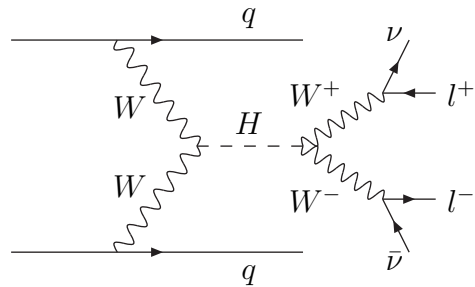

The Higgs at the LHC is produced predominantly via gluon-gluon
fusion. For Higgs masses such that $M_H>100\,\gev$ the second dominant
process is Vector Boson Fusion. The lowest order Feynman diagram of
the production of Higgs via VBF is depicted in Fig~\ref{fig:vbf_feynman}.
%
 The decay channel chosen is $H\rightarrow W^+W^-\rightarrow
e^{\pm}\mu^{\mp}\nu\overline{\nu},~e^{+}e^{-}\nu\overline{\nu},~\mu^{+}\mu^{-}\nu\overline{\nu}$.
These channels will also be referred to as $e\mu$, $ee$, and $\mu\mu$,
respectively.

These analyses were performed at the parton level and indicated that
this process could be the most powerful discovery mode at the LHC in
the range of the Higgs mass, $M_H$,
$115<M_H<200\,\gev$~\cite{pr_160_113004}.  These analyses were studied
specifically in the ATLAS environment using a fast simulation of the
detector~\cite{ATLFAST}.  Two traditional cut analyses, one for a
broad mass range and one optimized for a low-mass Higgs, were
developed and documented in references~\cite{UW-LowMass-VBF}
and~\cite{SciNote}.  We present results from previous studies without
systematic errors on the dominant $t\bar{t}$ background included.

In order to demonstrate the potential for multivariate algorithms, a
Neural Network analysis was performed~\cite{UW-nn-VBF}.  The approach
in the Neural Network study was to present a multivariate analysis
comparable to the cut analysis presented in~\cite{UW-LowMass-VBF}.
Thus, the analysis was restricted to kinematic variables which were
used or can be derived from the variables used in the cut analysis.

The variables used were:
\begin{itemize}

\item $\Delta \eta_{ll}$ -  the pseudorapidity difference between the
two leptons,

\item $\Delta \phi_{ll}$ -  the azimuthal angle difference between the
two leptons,

\item $M_{ll}$ -  the invariant mass of the two leptons,

\item $\Delta \eta_{jj}$ -  the pseudorapidity difference between the
two tagging jets,

\item $\Delta \phi_{jj}$ -  the azimuthal angle difference between the
two tagging jets,

\item $M_{jj}$ -  the invariant mass of the two tagging jets, and

\item $M_T$ -  the transverse mass.

\end{itemize}

\begin{figure*}
\begin{center}
\epsfig{file=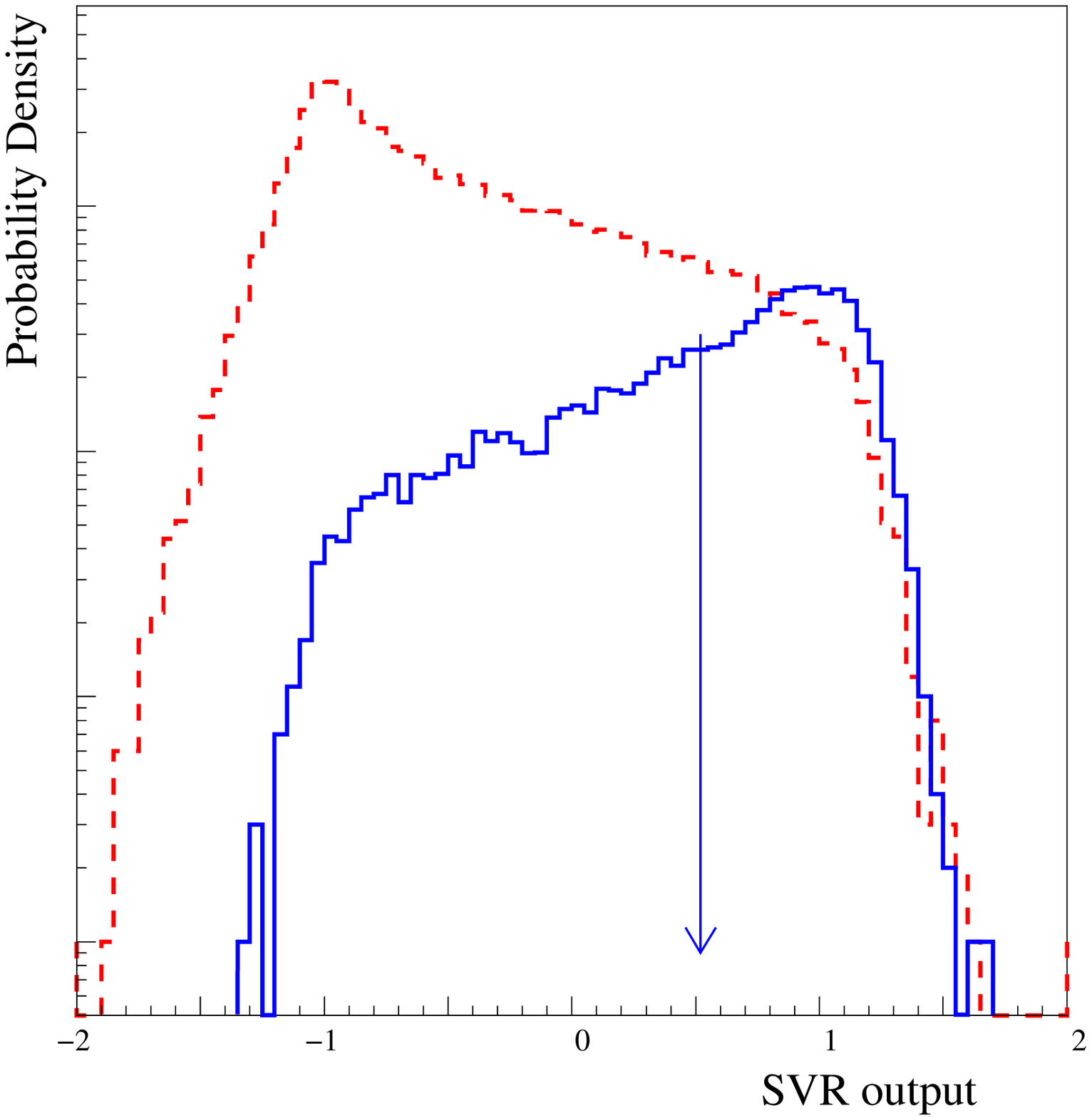, height= 2.5 in} \hspace{.4in}
\epsfig{file=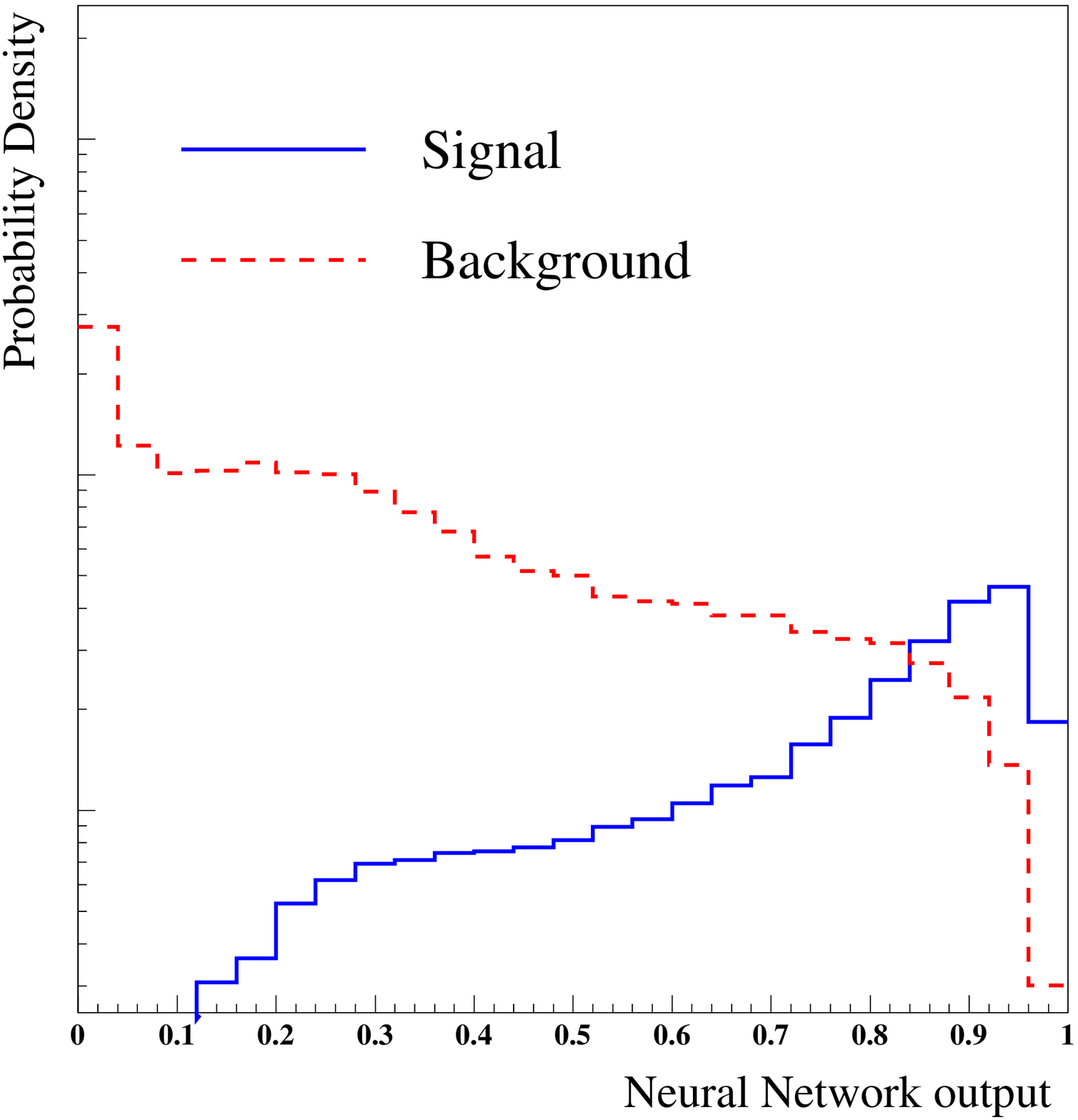, height= 2.5 in}
\end{center}
\caption{Support Vector Regression and Neural Network output
distributions for signal and background for 130 GeV Higgs boson in the
$e\mu$ channel.}
\label{fig:SVRandNN}
\end{figure*}

In total, three multivariate analyses were performed:
\begin{itemize}
  \item a Neural Network analysis using backpropagation with momentum,
  \item a Support Vector Regression analysis using Radial Basis
        Functions, and
  \item a Genetic Programming analysis using the software described in
        this Communication. 
\end{itemize}

The Neural Network (NN) analysis is well documented in reference
~\cite{UW-nn-VBF}.  The analysis were performed with both the Stutgart
Neural Network Simulator (SNNS)\footnote{SNNS can be found here:\\{\tt
www-ra.informatik.uni-tuebingen.de}} and MLPfit\footnote{MLPfit
can be found here:\\ {\tt cern.ch/\~{}schwind/MLPfit.html}} with a
7-10-10-1 architecture.

For the Support Vector Regression (SVR) analysis, the {\tt
BSVM-2.0}\footnote{BSVM can be found here: \\{\tt
www.csie.ntu.edu/\~{}cjlin/bsvm}} library was used.

The only parameters are the cost parameter, set to $C=1000$, and the
kernel function.  BSVM does not support weighted events, so an
``unweighted'' signal and background sample was used for training.

Because the trained machine only depends on a small subset of
``Support Vectors'', performance is fairly stable after only a
thousand or so training samples.  In this case, 2000 signal and 2000
background training events were used.

Both NN and SVR methods produce a function which characterizes the
signal-likeness of an event.  A separate procedure is used to find the
optimal cut on this function which optimizes the performance measure
(in this case the Poisson signficance, $\sigma_P$).
Fig.~\ref{fig:SVRandNN} shows the distribution of the SVR (left) and
NN (right) output values.  The optimal cut for the SVR technique is
shown as a vertical arrow.

Tab.~\ref{tab:MethodComparison} compares the Poisson significance,
$\sigma_P$, for a set of reference cuts, a set of cuts specifically
optimized for low-mass Higgs, Neural Networks, Genetic Programming,
and Support Vector Regression.  It is very pleasing to see that the
multivariate techniques achieve similar results.  Each of the methods
has its own set of advantages and disadvantages, but taken together
the methods are quite complementary.

\begin{table*}
\begin{center}
\begin{tabular}{l|ccccc}
{}           &  Ref. Cuts   &  low-$m_H$ Opt. Cuts   & NN & GP & SVR  \\ \hline
120 $ee$     &  0.87 &  1.25 & 1.72 & 1.66 & 1.44\\ 
120 $e\mu$   &  2.30 &  2.97 & 3.92 & 3.60 & 3.33\\ 
120 $\mu\mu$ &  1.16 &  1.71 & 2.28 & 2.26 & 2.08\\ 
Combined     &  2.97 &  3.91 & 4.98 & 4.57 & 4.26\\ \hline 
130 $e\mu$   &  4.94 &  6.14 & 7.55 & 7.22 & 6.59\\   

\end{tabular}
\end{center}
\caption{Expected significance for two cut analyses and three
multivariate analyses for different Higgs masses and final state
topologies.  Significance is expressed in terms of Gaussian Sigmas,
but calculated with Poisson statistics.}
\label{tab:MethodComparison}
\end{table*}

\section{Conclusions}

We have presented an implementation of a Genetic Programming system
specifically applied to the search for new particles.  In our approach
a group of individuals competes with respect to a user-defined
performance measure.  The genotype we have chosen consists of Boolean
conjunctions of simple arithmetic expressions of the input variables
required to lie in the interval $(-1,1)$.  Our implementation includes
an island model of parallelization and a recentering algorithm to
dramatically improve performance.  We have emphasized the importance
of the performance measure and decoupled fitness evaluation from
the optimization component of the algorithm.
We have touched on the relationship of Statistical Learning Theory and
VC dimension to the search for new particles and multivariate analysis
in general.  Finally, we have demonstrated that this method has
similar performance to Neural Networks (the de facto
multivariate analysis of High Energy Physics) and Support Vector
Regression.  We believe that this technique's most relevant advantages
are
\begin{itemize}
  \item the ability to provide a user-defined performance measure
  specifically suited to the problem at hand,
  \item the speed with which the resulting individual / cut can be evaluated,
  \item the fundamentally important ability to inspect the resulting
        cut, and
  \item the relatively low VC dimension which implies the method needs
  only a relatively small training sample.
\end{itemize}

\section{Acknowledgments}
This work was supported by a graduate research fellowship from the
National Science Foundation and US Department of Energy Grant
DE-FG0295-ER40896.
 


\bibliographystyle{unsrt}
\bibliography{vbf,bruce_cites,stats,PhysicsGP,nn}

\begin{thebibliography}{10}

\bibitem{koza:gp}
J.R. Koza.
\newblock {\em Genetic Programming: On the Programming of Computers by Means of
  Natural Selection}.
\newblock MIT Press, Cambridge, MA, 1992.

\bibitem{kishore:2000}
J.K.~Kishore et. al.
\newblock Application of genetic programming for multicategory pattern
  classification.
\newblock {\em IEEE Transactions on Evolutionary Computation}, 4 no.3, 2000.

\bibitem{Cranmer:Acta}
K.~Cranmer.
\newblock Multivariate analysis and the search for new particles.
\newblock {\em Acta Physica Polonica B}, 34:6049--6069, 2003.

\bibitem{Cranmer:2003vu}
K.~Cranmer.
\newblock Multivariate analysis from a statistical point of view.
\newblock In {\em PhyStat2003}, 2003.
\newblock physics/0310110.

\bibitem{Field:1997kt}
R.~D. Field and Y.~A. Kanev.
\newblock Using collider event topology in the search for the six-jet decay of
  top quark antiquark pairs.
\newblock {\em hep-ph/9801318}, 1997.

\bibitem{luke00two}
S.~Luke.
\newblock Two fast tree-creation algorithms for genetic programming.
\newblock {\em IEEE Transactions on Evolutionary Computation}, 2000.

\bibitem{david95parallel}
D.~Andre and J.R. Koza.
\newblock Parallel genetic programming on a network of transputers.
\newblock In Justinian~P. Rosca, editor, {\em Proceedings of the Workshop on
  Genetic Programming: From Theory to Real-World Applications}, pages 111--120,
  Tahoe City, California, USA, 9 1995.

\bibitem{Werbos}
P.J. Werbos.
\newblock {\em The Roots of Backpropagation}.
\newblock John Wiley \& Sons., New York, 1974.

\bibitem{PDP1}
D.E.~Rumelhart {\it et. al.}
\newblock {\em Parallel Distributed Processing Explorations in the
  Microstructure of Cognition}.
\newblock The MIT Press, Cambridge, 1986.

\bibitem{Cousins:1992qz}
R.D. Cousins and V.L. Highland.
\newblock Incorporating systematic uncertainties into an upper limit.
\newblock {\em Nucl. Instrum. Meth.}, A320:331--335, 1992.

\bibitem{Cranmer:2003vt}
K.~Cranmer.
\newblock Frequentist hypothesis testing with background uncertainty.
\newblock In {\em PhyStat2003}, 2003.
\newblock physics/0310108.

\bibitem{Linnemann:2003vw}
J.~T. Linnemann.
\newblock Measures of significance in {HEP} and astrophysics.
\newblock In {\em PhyStat2003}, 2003.
\newblock physics/0312059.

\bibitem{FinalLHWG:2003sz}
Search for the standard model {H}iggs boson at {LEP}.
\newblock {\em Phys. Lett.}, B565:61--75, 2003.

\bibitem{Punzi}
G.~Punzi.
\newblock Sensitivity of searches for new signals and its optimization.
\newblock In {\em PhyStat2003}, 2003.
\newblock physics/0308063.

\bibitem{Scott}
D.~Scott.
\newblock {\em Multivariate Density Estimation: Theory, Practice, and
  Visualization}.
\newblock John Wiley and Sons Inc., 1992.

\bibitem{Cranmer:2000du}
K.~Cranmer.
\newblock Kernel estimation in high-energy physics.
\newblock {\em Comput. Phys. Commun.}, 136:198--207, 2001.

\bibitem{Vapnik:1979}
V.~Vapnik.
\newblock Estimation of dependences based on empirical data.
\newblock {\em Nauka}, 1979.
\newblock in Russian.

\bibitem{Vapnik:1968}
V.~Vapnik and A.J. Chervonenkis.
\newblock The uniform convergence of frequencies of the appearance of events to
  their probabilities.
\newblock {\em Dokl. Akad. Nauk SSSR}, 1968.
\newblock in Russian.

\bibitem{Sontag}
E.~Sontag.
\newblock {VC} dimension of neural networks.
\newblock In C.M. Bishop, editor, {\em Neural Networks and Machine Learning},
  pages 69--95, Berlin, 1998. Springer-Verlag.

\bibitem{pr_160_113004}
D.~Rainwater and D.~Zeppenfeld.
\newblock Observing {$H\rightarrow W^{(\star)}W^{(\star)}\rightarrow
  e^{\pm}\mu^{\pm}\sla{p_{T}}$} in weak boson fusion with dual forward jet
  tagging at the {CERN LHC}.
\newblock D60:113004, 1999.

\bibitem{ATLFAST}
D.~Froidevaux E.~Richter-Was and L.~Poggioli.
\newblock Atlfast2.0 a fast simulation package for atlas.
\newblock ATLAS Internal Note ATL-PHYS-98-131.

\bibitem{UW-LowMass-VBF}
K.~Cranmer {\it et. al.}
\newblock Search for {H}iggs bosons decay {$H\rightarrow W^+W^- \rightarrow
  l^{+}l^{-}\sla{p_{T}}$} for {$115<M_H<130\,\gev$} using vector boson fusion.
\newblock ATLAS note ATL-PHYS-2003-002 (2002).

\bibitem{SciNote}
S.~Asai {\it et. al.}
\newblock Prospects for the search of a standard model {H}iggs boson in {ATLAS}
  using vector boson fusion.
\newblock to appear in EPJ.
\newblock ATLAS Scientific Note ATL-PHYS-2003-005 (2002).

\bibitem{UW-nn-VBF}
K.~Cranmer {\it et. al.}
\newblock Neural network based search for {H}iggs bosons decay
  {$H\rightarrow~W^+W^-~\rightarrow~l^{+}l^{-}\sla{p_{T}}$} for
  {${115<M_H<130\,\gev}$}.
\newblock ATLAS note ATL-PHYS-2003-007 (2002).

\end{thebibliography}

\end{document}